\pdfoutput=1
\documentclass[%
 reprint,
 amsmath,amssymb, 
 aps, 
]{revtex4-1}

\usepackage{graphicx}
\usepackage{dcolumn}
\usepackage{bm}

\begin{document}

\preprint{APS/123-QED}

\title{Heat Diffusion Imaging: In-Plane Thermal Conductivity Measurement of Thin Films in a Broad Temperature Range}

\author{T. Zhu}
\affiliation{Department of Electrical and Computer Engineering, Charlottesville, VA, 22904, USA}
\author{D. H. Olson} 
\affiliation{Department of Mechanical and Aerospace Engineering, University of Virginia, Charlottesville, VA, 22904, USA}
\author{P. E. Hopkins}
\affiliation{Department of Mechanical and Aerospace Engineering, University of Virginia, Charlottesville, VA, 22904, USA}
\affiliation {Department of Materials Science and Engineering, University of Virginia, Charlottesville, VA, 22904, USA}
\affiliation {Department of Physics, University of Virginia, Charlottesville, VA, 22904, USA}

\author{M. Zebarjadi}
\email{m.zebarjadi@virginia.edu.}
\affiliation{Department of Electrical and Computer Engineering, University of Virginia, Charlottesville, VA, 22904, USA} 
\affiliation {Department of Materials Science and Engineering, University of Virginia, Charlottesville, VA, 22904, USA 
}%

\date{\today}

\begin{abstract}

This work combines the principles of the heat spreader method and imaging capability of the thermoreflectance measurements to measure the in-plane thermal conductivity of thin-films without the requirement of film suspension or multiple thermometer deposition. We refer to this hybrid technique as \emph{heat diffusion imaging}. The thermoreflectance imaging system provides a temperature distribution map across the film surface. The in-plane thermal conductivity can be extracted from the temperature decay profile. By coupling the system with a cryostat, we were able to conduct measurements from 40\ K to 400\ K. Silicon thin film samples with and without periodic holes were measured and compared with in-plane time-domain thermoreflectance (TDTR) measurement and literature data as validation for heat diffusion imaging. 

\end{abstract}

\maketitle

\section{\label{sec:1}Introduction}

The continuous miniaturization of the electronic and optoelectronic devices results in increased energy dissipation densities and large local temperatures, which compromises the reliable operation of the electronics \cite{Pop2010} and calls for small scale thermal management solutions. Thin films are often used in integrated circuits, such as transistors and diodes \cite{Park2012}; in energy conversion technologies, such as LEDs and solar cells \cite{Green2007}; and as optical, electrical, thermal, or protective coating layers \cite{Martinu2000, Park2011}. Understanding heat transport in thin films is essential to address micro- to nanoscale heat management and energy conversion, and requires accurate measurement of the thermal conductivity, which can be challenging in practice. 

Measuring the in-plane thermal conductivity of thin film samples often requires film suspension \cite{Tai1988, Asheghi2002, Liu2004} and is demanding in terms of the micro-fabrication processes needed. In the case of thin films supported on a substrate, which is often the case in devices, the options are limited. If the samples are isotropic, a cross-plane measurement can be taken instead, using the 3$\omega$ method \cite{Cahill1990} or the time-domain thermoreflectance (TDTR) method \cite{Cahill2004}. In the case of anisotropic samples,
two established in-plane measurement techniques are the variable-linewidth 3$\omega$ method \cite{Ju1999, Kurabayashi1999} and the heat spreader method \cite{Jang2010, Dames2013}. Both methods require extensive fabrication work and can be difficult to implement. There has been an attempt to combine thermoreflectance imaging with finite element modeling for measurement of the in-plane thermal conductivity of Si thin films at room temperature \cite{Aubain2011}, but it requires intensive modeling and is considered as an indirect method to extract thermal conductivity. Similar combinations of thermoreflectance imaging and finite element modeling have been used by many groups to study heat transport in devices and extract various parameters, such as thermal boundary resistance between the layers and thermal conductivity of each layer of the device \cite{Bian2003, Zhang2006, Favaloro2014, Maize2014, Nadri2019}. 
While additional opportunities for measuring in-plane thermal conductivities can be provided using TDTR, typical measurements of the technique are utilized for the extraction of the cross-plane thermal conductivity of a material.

Here, we propose an approach that combines thermoreflectance (TR) imaging and the heat spreader method for characterizing in-plane thermal conductivity of thin films on substrate, and we  refer  to  this  method  as  heat  diffusion  imaging. A heater is deposited at one end of the film and subsequently Joule heated. When heat flow reaches the thin film of interest, the film, supported by a low thermal conductivity substrate, spreads the heat laterally. Therefore, it is possible to extract  its in-plane thermal conductivity by measuring the temperature decay profile along the film. This is known as the heat spreader method. Traditional heat spreader method requires patterning of a series of metallic lines along the sample. These metallic lines are used as resistance thermometers (thermistors) to detect the local temperatures. To measure temperature, one needs to pass current and measure resistance of the thermistors, hence, deposition of an insulating layer between a conductive sample and these thermometers is needed to avoid a current leak. This indicates that a limited number of data points (equal to the number of thermistors deposited) is available for thermal conductivity extraction, and inaccuracy exists because the temperature is not measured directly on sample. 

With a TR imaging system, the thermistor fabrication processes are no longer necessary. The temperature can be measured directly on the sample and continuously along the film with a spatial resolution on the order of 100\,nm. A temperature map of the sample surface is acquired based on how surface reflectivity changes with respect to temperature variations. With heat diffusion imaging, one only needs a metallic heater line deposited on top of the thin film to provide the initial heat flow by Joule heating.  

We select silicon thin films as the test bed for our heat diffusion imaging technique, since Si is one of the most studied materials and can serve as a good reference. We measure two samples: one is a plain Si thin film, and the other is a holey Si thin film with periodic holes spaced 100\,nm apart. Both films are 100\,nm thick. In-plane thermal conductivity of the holey Si sample is greatly reduced compared to that of the in-plane silicon film, as phonons with mean free paths longer than the neck size are suppressed \cite{Yu2010, Hao2016, Tang2010, Lim2016, Nomura2018}. Information on this reduction in thermal conductivity can shed light on the phonon interactions with boundaries and can be useful in finding low thermal conductivity films and, in particular, efficient thermoelectric materials \cite{Tang2010, Lim2016, Nomura2018}. Our measurement results are compared with those measured by in-plane TDTR on the same samples,
as well as data from literature.  

\begin{figure}[h]
\includegraphics[width=0.45\textwidth]{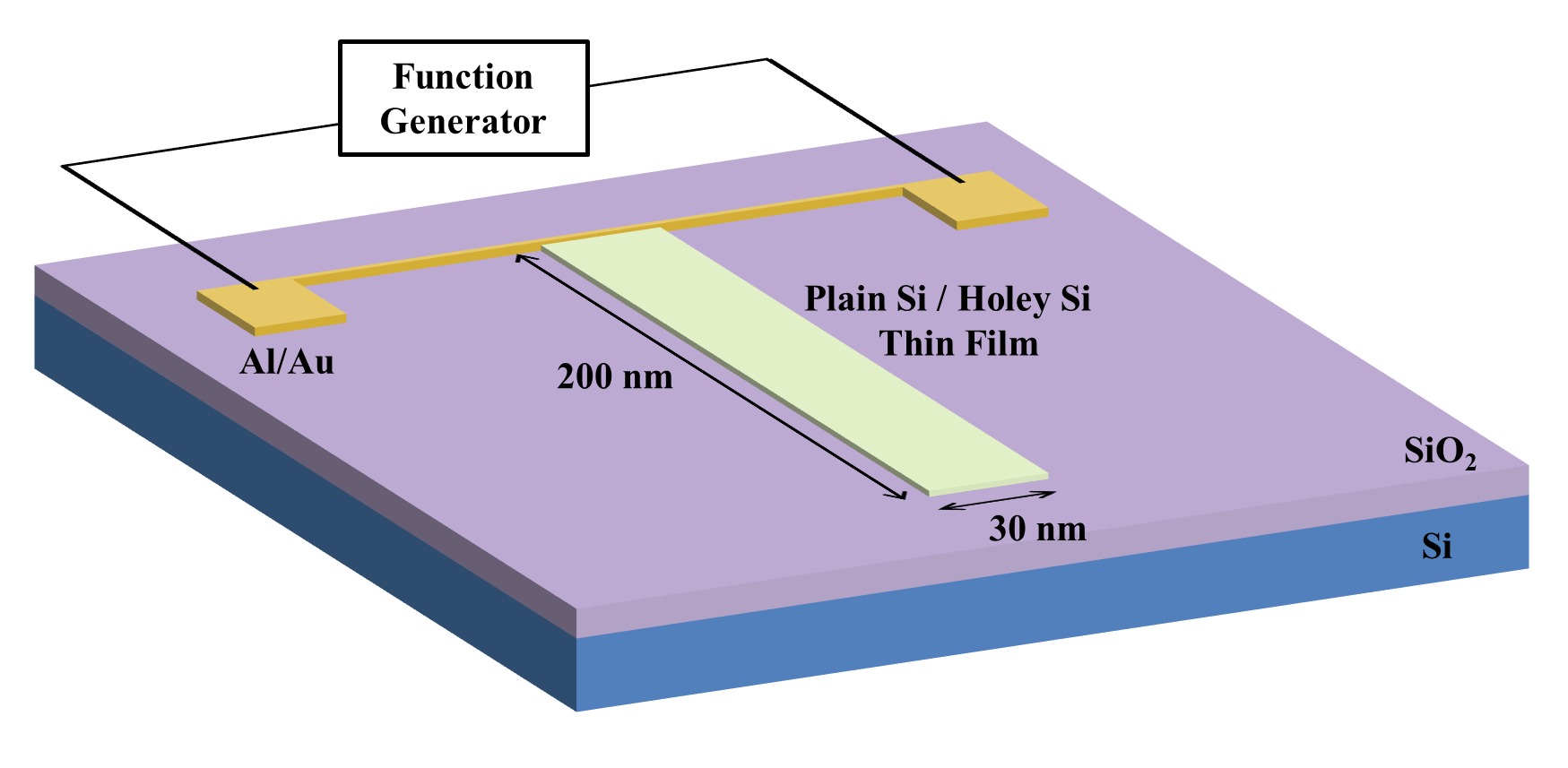}
\caption{\label{device_config} Schematic drawing of the device configuration for measurement. The 100\,nm thick plain or holey Si thin film of area 30$\times$200\,$\mu$m$^2$ sits on 2\,$\mu$m thick SiO$_2$ on bulk Si substrate. The metal electrodes at the two ends of the films are made of 1\,$\mu$m thick Al capped with 50\,nm thick Au. The dimensions are not to scale in this drawing. }
\end{figure}

\section{\label{sec:2}Device and Method}
\subsection{Silicon Thin Films}

An illustration of the  two silicon thin film samples is shown in Fig.\,\ref{device_config}. These devices are fabricated from a silicon on insulator (SOI) wafer and have a 100\ nm thick active Si layer of area 30$\times$200\,$\mu$m$^2$, sitting on a 2\,$\mu$m thick buried oxide (BOX). The holey Si device is boron-doped (3\,--\,10$\times10^{19}$\,cm$^{-3}$). The arrays of holes of 55\,nm in diameter are spaced with a center-to-center distance of 100\,nm or, in other words, a neck size of 45\,nm. The plain Si thin film is lightly doped (boron, $\sim\,10^{16}$\,cm$^{-3}$) and has the same dimensions as the holey silicon film but without the holes. The Al/Au contact (1\,$\mu$m/50\,nm) deposited at the end of the thin film is used as the heater for our measurements.

\subsection{Heat Diffusion Imaging} 

A thermoreflectance imaging 
system utilizes the fact that the surface reflectivity ($R$) of a material changes with temperature ($T$). Their relation is expressed as $\Delta T=\frac{1}{C_{TR}}\frac{\Delta R}{R}$, where $C_{TR}$ is the thermoreflectance coefficient. $C_{TR}$ mainly depends on the material surface, the ambient temperature, and the wavelength of the illumination. In order to find the absolute temperature on the sample surface, $C_{TR}$ has to be known and it can be calibrated for with controlled temperature changes. 

\begin{figure}[t]
    
    \includegraphics[width=0.4\textwidth]{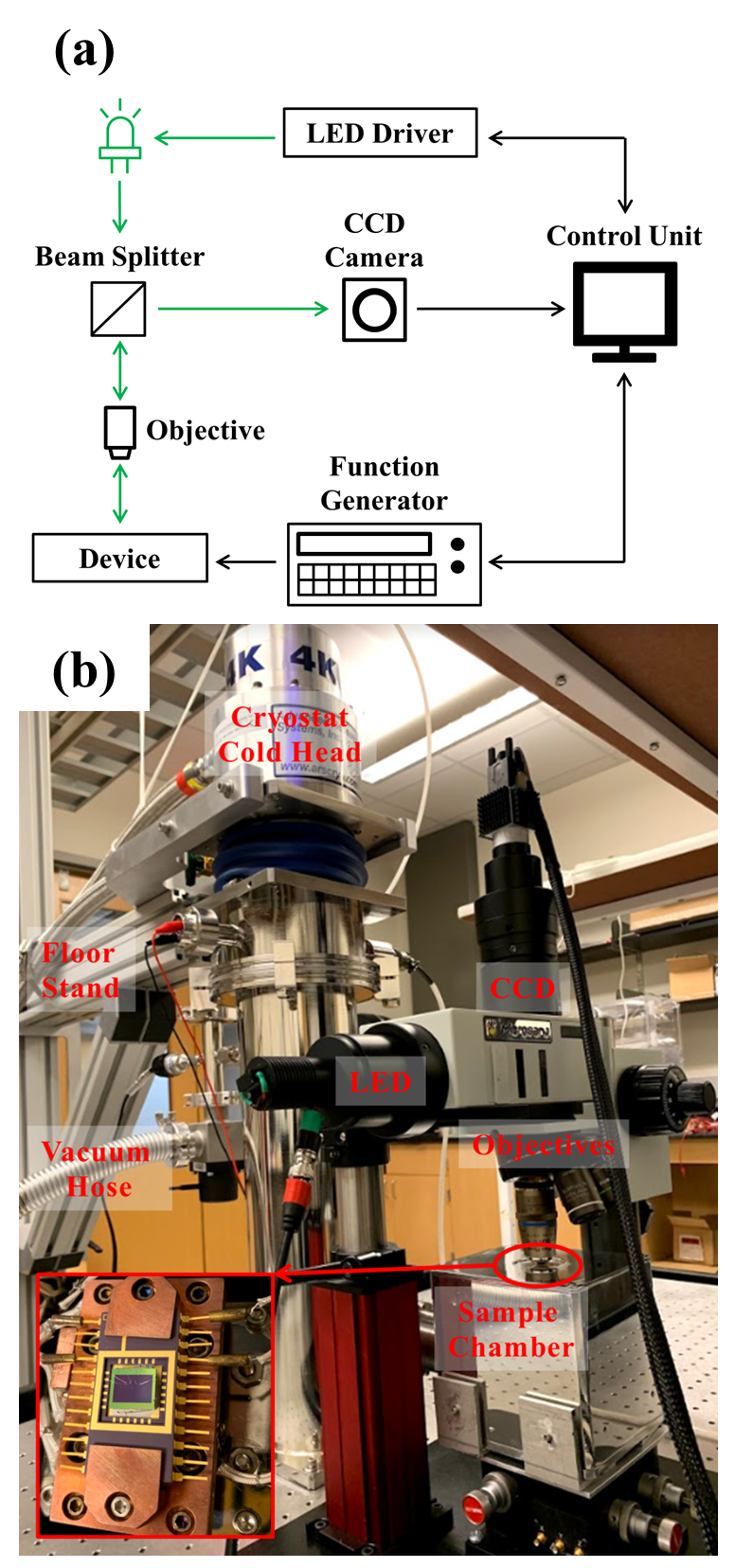}
    \caption{(a)\,The working principle of our thermoreflectance imaging system from Microsanj. (b)\,Our thermoreflectance imaging system coupled with a cyrostat. The inset shows how a sample is mounted to a DIP and then to the sample stage the inside of the sample chamber. }
    \label{fig_ars}
\end{figure}

A working diagram of our TR imaging system from Microsanj, LLC \cite{Christofferson2005} is shown in Fig.\,\ref{fig_ars}(a). The function generator generates a voltage pulse to excite the device under study, which is synchronized with a light pulse emitted from the LED light source (530\,nm green light in this case) by the control unit. A charge-coupled device (CCD) camera captures the changes in the reflected light off the sample when the LED is activated and deactivated to determine the temperature change $\Delta T$ as a function of time. The captured reflectivity information is averaged over many thermal excitation cycles to provide a complete temperature map of the area of interest with minimum noise. The magnification of the microscope objective can be selected based on sample size. The resolution of a temperature map is about 60\,nm per pixel under a 100$\times$ magnification objective and about 100\ nm per pixel under 60$\times$ magnification.  

The in-plane thermal conductivity can be extracted by analyzing how heat propagates along the thin films. When current passes though the metallic line at one end of the Si thin film, the generated heat from Joule heating results in a temperature gradient, which gradually decays with increasing distance away from the heater line. This temperature decay can be described by the classical fin equation \cite{Incropera2007} and is expected to be exponential in the lateral direction. Temperature, $T$, as a function of distance away from the heater, $x$, is: 
\begin{equation}
    T(x)-T(\infty)\propto e^{-\sqrt{h_i/\left(k_x d\right)} x} = e^{-\beta x}, \label{eq_1}
\end{equation}
where the parameter $\beta=\sqrt{h_i/\left(k_x d\right)}$ is defined for simplicity, 
$h_i$ is the cross-plane thermal conductance of the the underlying insulating layer, $k_x$ is the in-plane thermal conductivity of the thin film of interest and $d$ is its thickness. Dividing the cross-plane thermal conductivity of the insulator, $k_{i,z}$, by the layer thickness, $d_i$, gives us $h_i$, i.e., $h_i=k_{i,z}/d_i$. $k_x$ can be extracted from the temperature profile as long as the thin film thickness and the substrate parameters are known. This theoretical model has been discussed extensively in the heat spreader method \cite{Dames2013} and has been verified for Si thin films on SiO$_2$/Si substrate \cite{Asheghi1998} and for few layer graphene on SiO$_2$ \cite{Jang2010}.

Fig.\,\ref{fig_ars}(b) shows the entire setup. The cryostat from Advanced Research Systems (ARS) has been specially engineered so that the sample chamber has an optical window on top for imaging. The window is made of 1\,mm thick N-BK7 glass with Vis 0$^{\circ}$ anti-reflective coating from Edmund Optics, which provides $>99.5\%$ transmissivity for wavelength between 425 and 675\,nm and has a transformation temperature above 800\,K.  The thickness of the window glass needs to be corrected for by the correction collar of the objective, in order to avoid blurry images caused by changes in the refractive index.We used the 60$\times$ magnification objective from Nikon which can correct for cover glass thickness up to 1.3\,mm and offers a relatively long working distance up to 2.6\,mm. The sample is wire-bonded to a ceramic dual in-line package (DIP) for electrical connections inside the chamber and is shown in the inset of Fig.\,\ref{fig_ars}(b). A heater is embedded in the sample stage to set the temperature. Vibrations caused by the vacuum pump and the helium compressor have been minimized by using low-vibration design from ARS, placing the setup on an optical table, having a floor stand and sandbags to support the cold head, and a stand to support the TR imaging camera system.  

\subsection{In-Plane Time-Domain Thermoreflectance}

We deposit a nominally 80\,nm thick Al film on an identical plain Si device for the implementation of time-domain thermoreflectance (TDTR) to extract its in-plane thermal conductivity. We implement a two-tint TDTR configuration, whereby the output of a 80\,MHz oscillator centered at 808.5\,nm is spectrally separated into pump and probe paths. The pump path is electro-optically modulated, and creates a frequency dependent heating event at the sample surface. The probe is mechanically delayed in time and monitors the thermoreflectance at the sample surface, creating a cooling curve that is compared to the heat diffusion equation. Additional details regarding the two-tint experimental configuration \cite{Kang2008}, as well as the analyses associated with TDTR \cite{Cahill2004, Hopkins2010, Schmidt2008}, can be found in the literature. We mount the specimen into a liquid helium cryostat (JANIS ST-100) for measurements at 80, 124, 193, and 294 K. 

For the determination of the in-plane thermal conductivity of the Si membrane, we perform measurements in two configurations, following the guidelines of Jiang \emph{et al.} \cite{Jiang2017}. First, we perform TDTR measurements using a 10$\times$ objective (effective $1/e^2$ pump/probe radii = 12.3\,$\mu$m) and moderately high modulation frequencies (8.4\,MHz) on the Al/SiO$_2$ region of the device. This allows us to extract the thermal conductivity of the oxide layer while being negligibly sensitive to the in-plane thermal conductivity of the Al transducer. Following, we perform measurements using a 20$\times$ objective (effective $1/e^2$ pump/probe radii = 6.5\,$\mu$m) and low modulation frequencies (500\,kHz) on the Al/SiO$_2$ region of the device for the extraction of the in-plane thermal conductivity of the Al transducer, while fixing the thermal conductivity of the underlying SiO$_2$. Finally, measurements in this configuration are performed on the Al/Si/SiO$_2$ stack for the determination of the in-plane thermal conductivity of the Si membrane, where we input the Al and SiO$_2$ thermal conductivities as known parameters derived from measurements on the Al/SiO$_2$ region of the device. For all measurements, we assume literature values for the volumetric heat capacities of Al, SiO2, and Si \cite{Lide2005}.  The error in the reported TDTR values includes uncertainties of 2.5\% in the volumetric heat capacities of the Al transducer, Si membrane, and SiO$_2$ layer, as well as variations in the extracted Al thermal conductivity, typically on the order of 15-20\%.

\section{\label{sec:3}Results and Discussion}

\begin{figure*}[t]
    \includegraphics[width=0.9\textwidth]{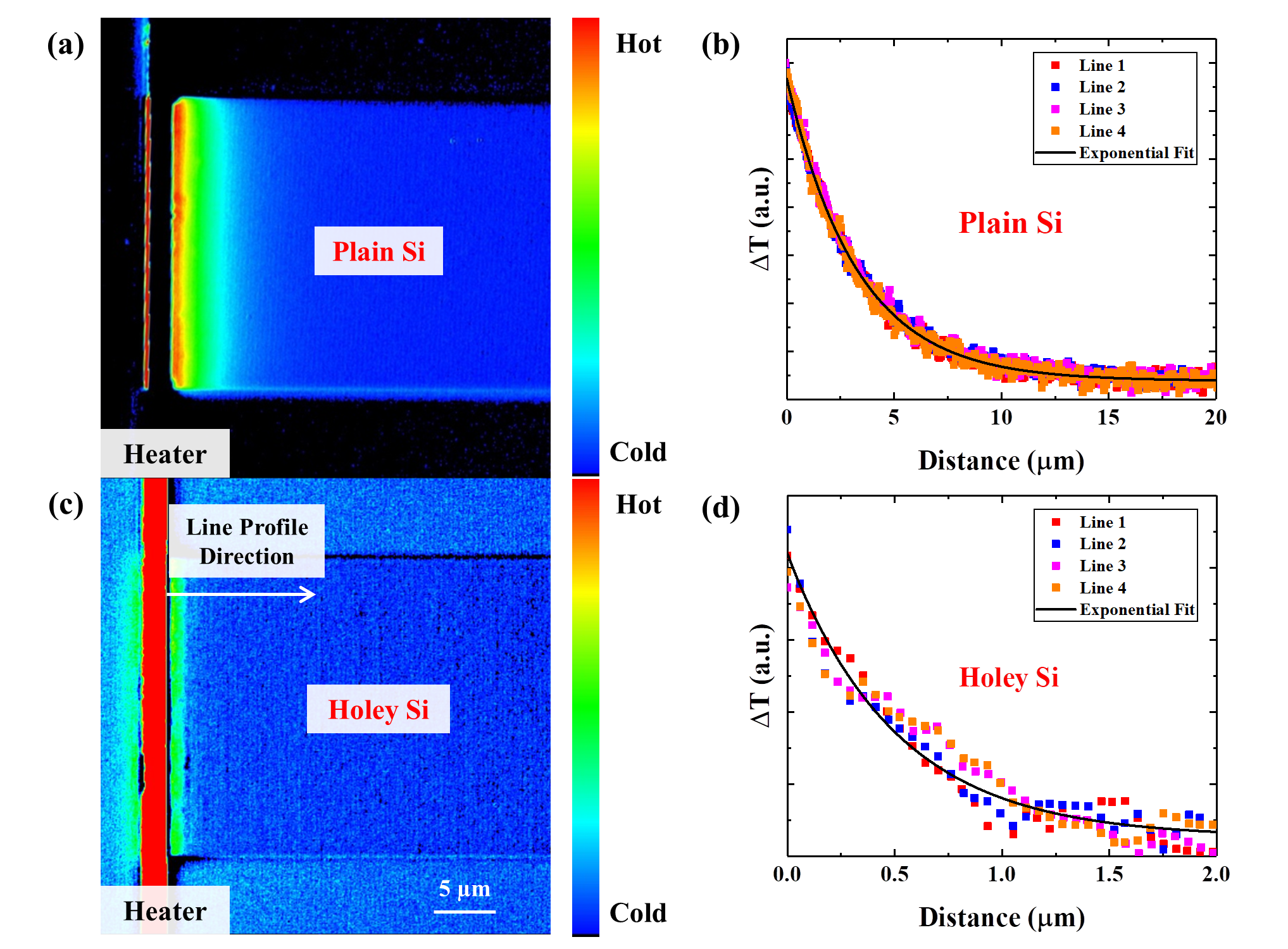}
    \caption{\label{T_map} Example temperature maps of (a) the plain Si sample and (c) the holey Si sample under 100$\times$ magnification at room temperature. Note that the heater in (a) is shown in dark color as its thermoreflectance coefficient has an opposite sign to the coefficient assumed. Typically more than ten temperature line profiles are taken from each temperature map for averaging. A few representative temperature decay curves and their corresponding exponential fitting curves are shown for (b) the plain Si and (d) the holey Si thin films. The same thermoreflectance coeffcient is assumed for both maps and an arbitrary unit for temperature is adopted.}
\end{figure*}

Pulses of 5\,ms in duration, up to 5\,V in voltage and 20\% duty cycle were applied to the heater line at one end of the thin film to generate Joule heating. An example of the temperature decay map of the plain Si thin film in steady state, averaged over a few hundred thermal excitation cycles and taken under 100$\times$ magnification, is shown in Fig.\,\ref{T_map}(a). A representative map of the holey Si sample is provided by its side in Fig.\,\ref{T_map}(c) for comparison. 
The same thermoreflectance coefficient, $C_{TR}$, is used for all temperature maps across all materials, which means that an arbitrary unit is adopted for temperature. Calibrating for $C_{TR}$ to get the absolute temperatures on the thin films is not necessary here, because the parameter $\beta$ is determined by an exponential fit of the temperature profile and depends only on how fast the temperature drops. Nevertheless, since $C_{TR}$ of Au for our green light wavelength is known \cite{Favaloro2015}, the temperature rise of the heater (less than 50\,K) was closely monitored throughout measurements to ensure that it is reasonable to assume a single thermoreflectance coefficient. The distance that the temperature gradient extends over is much shorter on the holey Si thin film than on the non-holey Si one, because of a greatly suppressed thermal conductivity.

Knowing that the Si thin film thickness $d=100$\,nm and that the insulating SiO$_2$ layer thickness $d_{i}=2\,\mu$m, the in-plane thermal conductivity calculation is reduced to 
\begin{equation}
k_x=\frac{h_i}{ \beta^2d}=5\times10^{12}\times\frac{ k_{i,z}}{\beta^{2}}. 
\end{equation}
We use the recommended values of fused SiO$_2$ \cite{Touloukian1970} for $k_{i,z}$ in our calculations, as it has been shown that the temperature dependence of the thermal conductivity of micrometers-thick SiO$_2$ matches well with the bulk values \cite{Asheghi1998}. $h_i$ is typically on the order of $10^5-10^6$\,W/(m$^2$K), which means that the air convection to the surroundings \cite{Incropera2007} and the thermal contact resistance between the Si layer and the SiO$_2$ layer \cite{Mahajan2011, Chen2012, Lampin2012} are negligible in comparison. An exponential fit to the temperature line profile taken along the thin film yields $\beta$ and then $k_{x}$ is obtained. A few representative temperature profiles of both films and their corresponding exponential fitting curves at room temperature are shown in Fig.\,\ref{T_map}(b) and (d). As expected, $\beta$ is smaller for plain Si based on a more gradual temperature decay, which results in a larger $k_x$. In order to reduce the uncertainty, for each measurement, multiple temperature line profiles that start from the edge of the heater and continue along the length of the thin films have been taken into account and an averaged thermal conductivity is reported.

Temperature dependent data were acquired by repeating the measurements at different sample stage temperatures. 
The highest magnification offered for objective with correction collar and adequate working distance is 60$\times$, so the temperature map resolution is reduced to about 100\,nm per pixel at temperatures other than room temperature. Limited by the working distance of this microscope objective, the mounted sample had to be brought to within 1\,mm of the optical window. This led to a discrepancy between the sample stage temperature and the sample surface temperature. This temperature difference was measured using a silicon diode temperature sensor inside the cryostat, by mounting the sensor onto the DIP, where the sample was mounted. The error in the measured sample temperature is estimated to be within 1\,K and has been taken into account in the thermal conductivity calculation.   

\begin{figure}[t]
    \includegraphics[    width=0.45\textwidth]{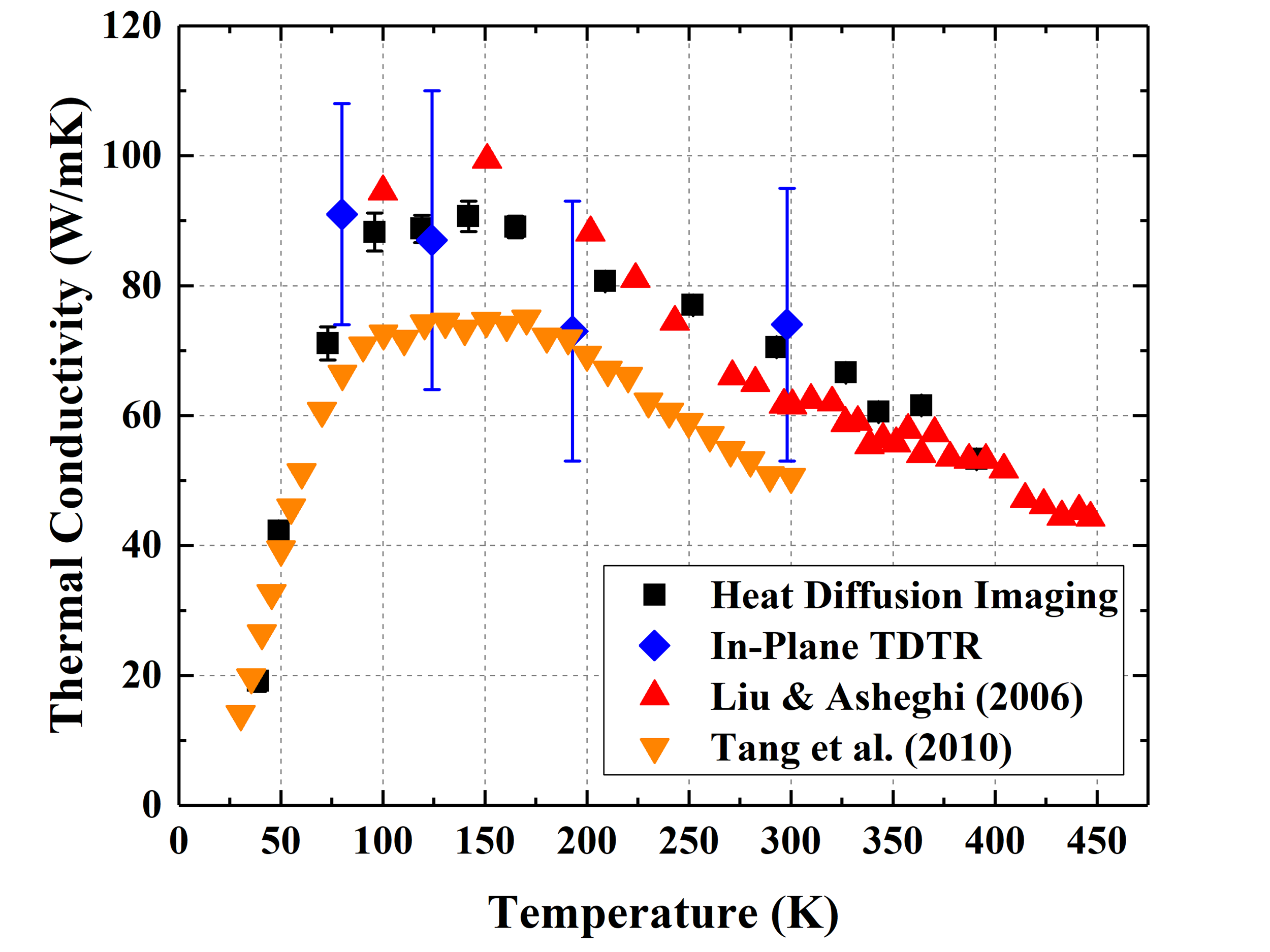}
    \caption{\label{Si_k} Temperature dependence of the in-plane thermal conductivity of the plain Si thin film measured by heat diffusion imaging, compared with in-plane TDTR measurement and data from literature  on samples of the same thickness \cite{Liu2006, Tang2010}. }
\end{figure}

The temperature dependence of the in-plane thermal conductivity of the plain Si film is plotted in Fig.\,\ref{Si_k} and is compared with data from literature \cite{Tang2010, Liu2006}. Phonon-boundary scattering dominates phonon-impurity scattering in  micrometer-thick plain Si thin films with carrier concentration less than 10$^{17}$\,cm$^{-3}$, and their thermal conductivities are nearly identical against temperature, independent of the exact concentrations \cite{Asheghi2002}. Therefore, it is justifiable to compare the results of our 100\,nm thick, slightly-doped (10$^{16}$\,cm$^{-3}$) plain Si thin film with thermal conductivities of intrinsic silicon thin films of the same thickness \cite{Liu2006, Tang2010}. Our values agree with data from literature in the temperature range from 40\,K to 400\,K with reasonable errors. While the uncertainty in our TDTR measurements are large, there is generally good agreement found with values extracted from heat diffusion imaging. 
These comparisons show the feasibility and reliability of the heat diffusion imaging method.  

\begin{figure}[t]
    \includegraphics[
    width=0.45\textwidth]{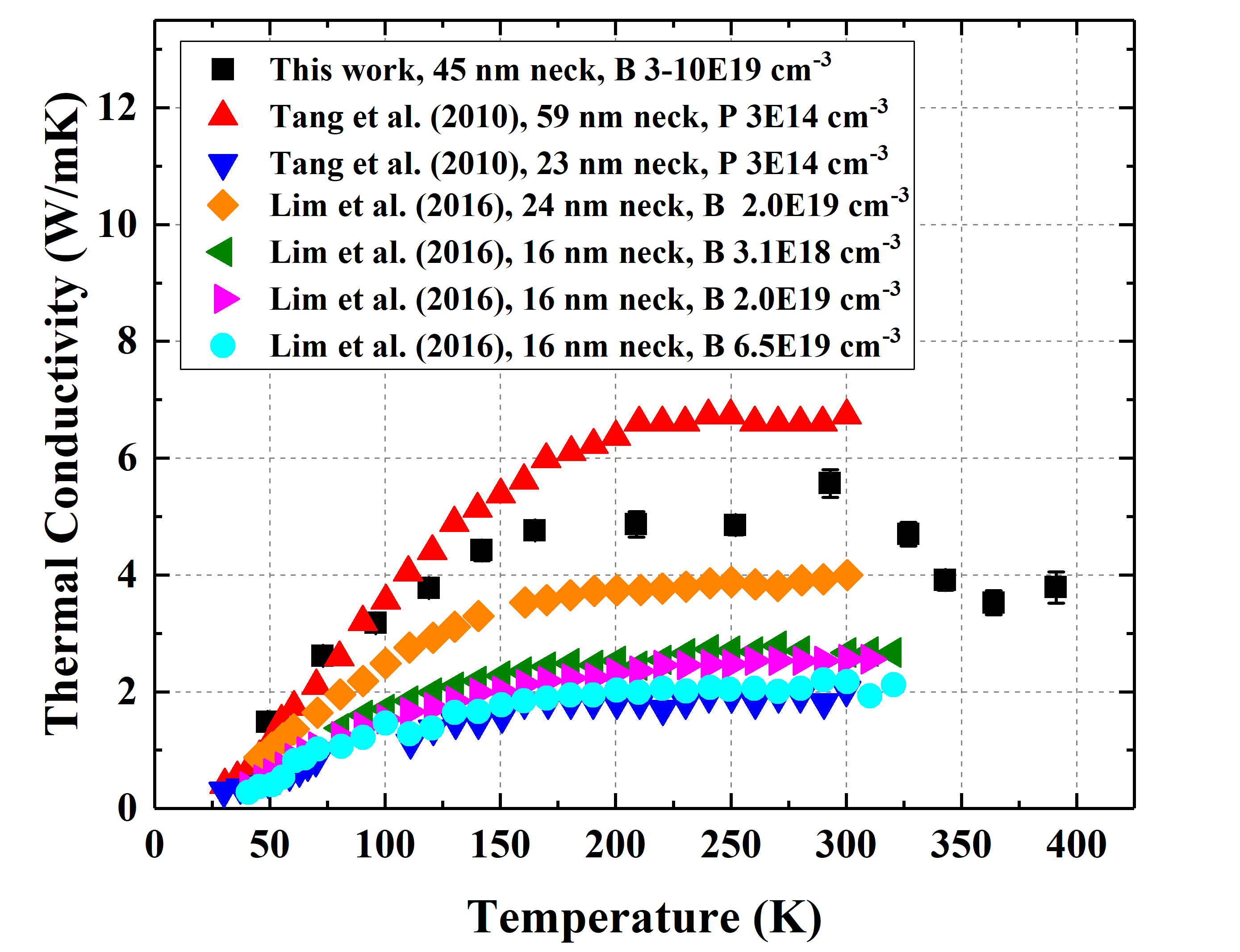}
    \caption{\label{HS_k} Temperature dependence of the in-plane thermal conductivity of the holey Si thin film measured by heat diffusion imaging, compared with holey Si thin films with the same thickness but different neck sizes from literature \cite{Tang2010, Lim2016}. The carrier concentration of each sample is noted in the legend, with B for boron-doped and P for phosphorus doped.  }
\end{figure}

Data for the holey Si sample are presented in Fig.\,\ref{HS_k}. The values are much lower than those of the plain Si sample, as the higher doping concentrations and periodic holes introduce more frequent phonon scattering events. We were not able to perform TDTR measurements on the in-plane thermal conductivity of the holey Si film due to its low thermal conductivity. As the neck size between holes is the major deciding factor in the suppression of the in-plane thermal conductivity \cite{Yu2010, Hao2016, Tang2010, Lim2016, Nomura2018}, we compared our results with holey Si samples of the same thickness but different neck sizes \cite{Tang2010, Lim2016}. Our values for 45\,nm neck size are lower than the data for 59\,nm neck size and higher than those with smaller neck sizes \cite{Tang2010, Lim2016}, as expected.

\section{\label{sec:4}Conclusion}

We have measured the in-plane thermal conductivity of holey Si and plain Si thin films from 40\,K to 400\,K using heat diffusion imaging, which combines thermoreflectance imaging with the heat spreader method. By comparing the obtained data for plain Si with values measured by in-plane TDTR and from literature, we have demonstrated the reliability of our method. The Heat diffusion imaging can be implemented to measure the thermal conductivity of supported thin films with minimum required micro-fabrication processes and is of great importance in the study of heat transport in thin films.

\begin{acknowledgments}

This work was supported by NSF grant number 1653268 and was supported in part under a MURI program through the Office of Naval Research, Grant No. N00014-18-1-2429. D. H. Olson is grateful for funding from the National Defense Science and Engineering Graduate (NDSEG) Fellowship. The authors would like to thank Prof. J.E. Bowers for providing thin-film samples. 

\end{acknowledgments}

\nocite{*}
\bibliography{references}

\begin{thebibliography}{39}%
\makeatletter
\providecommand \@ifxundefined [1]{%
 \@ifx{#1\undefined}
}%
\providecommand \@ifnum [1]{%
 \ifnum #1\expandafter \@firstoftwo
 \else \expandafter \@secondoftwo
 \fi
}%
\providecommand \@ifx [1]{%
 \ifx #1\expandafter \@firstoftwo
 \else \expandafter \@secondoftwo
 \fi
}%
\providecommand \natexlab [1]{#1}%
\providecommand \enquote  [1]{``#1''}%
\providecommand \bibnamefont  [1]{#1}%
\providecommand \bibfnamefont [1]{#1}%
\providecommand \citenamefont [1]{#1}%
\providecommand \href@noop [0]{\@secondoftwo}%
\providecommand \href [0]{\begingroup \@sanitize@url \@href}%
\providecommand \@href[1]{\@@startlink{#1}\@@href}%
\providecommand \@@href[1]{\endgroup#1\@@endlink}%
\providecommand \@sanitize@url [0]{\catcode `\\12\catcode `\$12\catcode
  `\&12\catcode `\#12\catcode `\^12\catcode `\_12\catcode `\%12\relax}%
\providecommand \@@startlink[1]{}%
\providecommand \@@endlink[0]{}%
\providecommand \url  [0]{\begingroup\@sanitize@url \@url }%
\providecommand \@url [1]{\endgroup\@href {#1}{\urlprefix }}%
\providecommand \urlprefix  [0]{URL }%
\providecommand \Eprint [0]{\href }%
\providecommand \doibase [0]{http://dx.doi.org/}%
\providecommand \selectlanguage [0]{\@gobble}%
\providecommand \bibinfo  [0]{\@secondoftwo}%
\providecommand \bibfield  [0]{\@secondoftwo}%
\providecommand \translation [1]{[#1]}%
\providecommand \BibitemOpen [0]{}%
\providecommand \bibitemStop [0]{}%
\providecommand \bibitemNoStop [0]{.\EOS\space}%
\providecommand \EOS [0]{\spacefactor3000\relax}%
\providecommand \BibitemShut  [1]{\csname bibitem#1\endcsname}%
\let\auto@bib@innerbib\@empty
\bibitem [{\citenamefont {Pop}(2010)}]{Pop2010}%
  \BibitemOpen
  \bibfield  {author} {\bibinfo {author} {\bibfnamefont {E.}~\bibnamefont
  {Pop}},\ }\href {\doibase 10.1007/s12274-010-1019-z} {\bibfield  {journal}
  {\bibinfo  {journal} {Nano Research}\ }\textbf {\bibinfo {volume} {3}},\
  \bibinfo {pages} {147} (\bibinfo {year} {2010})}\BibitemShut {NoStop}%
\bibitem [{\citenamefont {Park}\ \emph {et~al.}(2012)\citenamefont {Park},
  \citenamefont {Maeng}, \citenamefont {Kim},\ and\ \citenamefont
  {Park}}]{Park2012}%
  \BibitemOpen
  \bibfield  {author} {\bibinfo {author} {\bibfnamefont {J.~S.}\ \bibnamefont
  {Park}}, \bibinfo {author} {\bibfnamefont {W.-J.}\ \bibnamefont {Maeng}},
  \bibinfo {author} {\bibfnamefont {H.-S.}\ \bibnamefont {Kim}}, \ and\
  \bibinfo {author} {\bibfnamefont {J.-S.}\ \bibnamefont {Park}},\ }\href
  {\doibase 10.1016/j.tsf.2011.07.018} {\bibfield  {journal} {\bibinfo
  {journal} {Thin Solid Films}\ }\textbf {\bibinfo {volume} {520}},\ \bibinfo
  {pages} {1679} (\bibinfo {year} {2012})}\BibitemShut {NoStop}%
\bibitem [{\citenamefont {Green}(2007)}]{Green2007}%
  \BibitemOpen
  \bibfield  {author} {\bibinfo {author} {\bibfnamefont {M.~A.}\ \bibnamefont
  {Green}},\ }\href {\doibase 10.1007/s10854-007-9177-9} {\bibfield  {journal}
  {\bibinfo  {journal} {Journal of Materials Science: Materials in
  Electronics}\ }\textbf {\bibinfo {volume} {18}},\ \bibinfo {pages} {15}
  (\bibinfo {year} {2007})}\BibitemShut {NoStop}%
\bibitem [{\citenamefont {Martinu}\ and\ \citenamefont
  {Poitras}(2000)}]{Martinu2000}%
  \BibitemOpen
  \bibfield  {author} {\bibinfo {author} {\bibfnamefont {L.}~\bibnamefont
  {Martinu}}\ and\ \bibinfo {author} {\bibfnamefont {D.}~\bibnamefont
  {Poitras}},\ }\href {\doibase 10.1116/1.1314395} {\bibfield  {journal}
  {\bibinfo  {journal} {Journal of Vacuum Science {\&} Technology A: Vacuum,
  Surfaces, and Films}\ }\textbf {\bibinfo {volume} {18}},\ \bibinfo {pages}
  {2619} (\bibinfo {year} {2000})}\BibitemShut {NoStop}%
\bibitem [{\citenamefont {Park}\ \emph {et~al.}(2011)\citenamefont {Park},
  \citenamefont {Chae}, \citenamefont {Chung},\ and\ \citenamefont
  {Lee}}]{Park2011}%
  \BibitemOpen
  \bibfield  {author} {\bibinfo {author} {\bibfnamefont {J.-S.}\ \bibnamefont
  {Park}}, \bibinfo {author} {\bibfnamefont {H.}~\bibnamefont {Chae}}, \bibinfo
  {author} {\bibfnamefont {H.~K.}\ \bibnamefont {Chung}}, \ and\ \bibinfo
  {author} {\bibfnamefont {S.~I.}\ \bibnamefont {Lee}},\ }\href {\doibase
  10.1088/0268-1242/26/3/034001} {\bibfield  {journal} {\bibinfo  {journal}
  {Semiconductor Science and Technology}\ }\textbf {\bibinfo {volume} {26}},\
  \bibinfo {pages} {034001} (\bibinfo {year} {2011})}\BibitemShut {NoStop}%
\bibitem [{\citenamefont {Tai}\ \emph {et~al.}(1988)\citenamefont {Tai},
  \citenamefont {Mastrangelo},\ and\ \citenamefont {Muller}}]{Tai1988}%
  \BibitemOpen
  \bibfield  {author} {\bibinfo {author} {\bibfnamefont {Y.~C.}\ \bibnamefont
  {Tai}}, \bibinfo {author} {\bibfnamefont {C.~H.}\ \bibnamefont
  {Mastrangelo}}, \ and\ \bibinfo {author} {\bibfnamefont {R.~S.}\ \bibnamefont
  {Muller}},\ }\href {\doibase 10.1063/1.339924} {\bibfield  {journal}
  {\bibinfo  {journal} {Journal of Applied Physics}\ }\textbf {\bibinfo
  {volume} {63}},\ \bibinfo {pages} {1442} (\bibinfo {year}
  {1988})}\BibitemShut {NoStop}%
\bibitem [{\citenamefont {Asheghi}\ \emph {et~al.}(2002)\citenamefont
  {Asheghi}, \citenamefont {Kurabayashi}, \citenamefont {Kasnavi},\ and\
  \citenamefont {Goodson}}]{Asheghi2002}%
  \BibitemOpen
  \bibfield  {author} {\bibinfo {author} {\bibfnamefont {M.}~\bibnamefont
  {Asheghi}}, \bibinfo {author} {\bibfnamefont {K.}~\bibnamefont
  {Kurabayashi}}, \bibinfo {author} {\bibfnamefont {R.}~\bibnamefont
  {Kasnavi}}, \ and\ \bibinfo {author} {\bibfnamefont {K.~E.}\ \bibnamefont
  {Goodson}},\ }\href {\doibase 10.1063/1.1458057} {\bibfield  {journal}
  {\bibinfo  {journal} {Journal of Applied Physics}\ }\textbf {\bibinfo
  {volume} {91}},\ \bibinfo {pages} {5079} (\bibinfo {year}
  {2002})}\BibitemShut {NoStop}%
\bibitem [{\citenamefont {Liu}\ and\ \citenamefont {Asheghi}(2004)}]{Liu2004}%
  \BibitemOpen
  \bibfield  {author} {\bibinfo {author} {\bibfnamefont {W.}~\bibnamefont
  {Liu}}\ and\ \bibinfo {author} {\bibfnamefont {M.}~\bibnamefont {Asheghi}},\
  }\href {\doibase 10.1063/1.1741039} {\bibfield  {journal} {\bibinfo
  {journal} {Applied Physics Letters}\ }\textbf {\bibinfo {volume} {84}},\
  \bibinfo {pages} {3819} (\bibinfo {year} {2004})}\BibitemShut {NoStop}%
\bibitem [{\citenamefont {Cahill}(1990)}]{Cahill1990}%
  \BibitemOpen
  \bibfield  {author} {\bibinfo {author} {\bibfnamefont {D.~G.}\ \bibnamefont
  {Cahill}},\ }\href {\doibase 10.1063/1.1141498} {\bibfield  {journal}
  {\bibinfo  {journal} {Review of Scientific Instruments}\ }\textbf {\bibinfo
  {volume} {61}},\ \bibinfo {pages} {802} (\bibinfo {year} {1990})}\BibitemShut
  {NoStop}%
\bibitem [{\citenamefont {Cahill}(2004)}]{Cahill2004}%
  \BibitemOpen
  \bibfield  {author} {\bibinfo {author} {\bibfnamefont {D.~G.}\ \bibnamefont
  {Cahill}},\ }\href {\doibase 10.1063/1.1819431} {\bibfield  {journal}
  {\bibinfo  {journal} {Review of Scientific Instruments}\ }\textbf {\bibinfo
  {volume} {75}},\ \bibinfo {pages} {5119} (\bibinfo {year}
  {2004})}\BibitemShut {NoStop}%
\bibitem [{\citenamefont {Ju}\ \emph {et~al.}(1999)\citenamefont {Ju},
  \citenamefont {Kurabayashi},\ and\ \citenamefont {Goodson}}]{Ju1999}%
  \BibitemOpen
  \bibfield  {author} {\bibinfo {author} {\bibfnamefont {Y.}~\bibnamefont
  {Ju}}, \bibinfo {author} {\bibfnamefont {K.}~\bibnamefont {Kurabayashi}}, \
  and\ \bibinfo {author} {\bibfnamefont {K.}~\bibnamefont {Goodson}},\ }\href
  {\doibase 10.1016/S0040-6090(98)01328-5} {\bibfield  {journal} {\bibinfo
  {journal} {Thin Solid Films}\ }\textbf {\bibinfo {volume} {339}},\ \bibinfo
  {pages} {160} (\bibinfo {year} {1999})}\BibitemShut {NoStop}%
\bibitem [{\citenamefont {Kurabayashi}\ \emph {et~al.}(1999)\citenamefont
  {Kurabayashi}, \citenamefont {Asheghi}, \citenamefont {Touzelbaev},\ and\
  \citenamefont {Goodson}}]{Kurabayashi1999}%
  \BibitemOpen
  \bibfield  {author} {\bibinfo {author} {\bibfnamefont {K.}~\bibnamefont
  {Kurabayashi}}, \bibinfo {author} {\bibfnamefont {M.}~\bibnamefont
  {Asheghi}}, \bibinfo {author} {\bibfnamefont {M.}~\bibnamefont {Touzelbaev}},
  \ and\ \bibinfo {author} {\bibfnamefont {K.}~\bibnamefont {Goodson}},\ }\href
  {\doibase 10.1109/84.767114} {\bibfield  {journal} {\bibinfo  {journal}
  {Journal of Microelectromechanical Systems}\ }\textbf {\bibinfo {volume}
  {8}},\ \bibinfo {pages} {180} (\bibinfo {year} {1999})}\BibitemShut {NoStop}%
\bibitem [{\citenamefont {Jang}\ \emph {et~al.}(2010)\citenamefont {Jang},
  \citenamefont {Chen}, \citenamefont {Bao}, \citenamefont {Lau},\ and\
  \citenamefont {Dames}}]{Jang2010}%
  \BibitemOpen
  \bibfield  {author} {\bibinfo {author} {\bibfnamefont {W.}~\bibnamefont
  {Jang}}, \bibinfo {author} {\bibfnamefont {Z.}~\bibnamefont {Chen}}, \bibinfo
  {author} {\bibfnamefont {W.}~\bibnamefont {Bao}}, \bibinfo {author}
  {\bibfnamefont {C.~N.}\ \bibnamefont {Lau}}, \ and\ \bibinfo {author}
  {\bibfnamefont {C.}~\bibnamefont {Dames}},\ }\href {\doibase
  10.1021/nl101613u} {\bibfield  {journal} {\bibinfo  {journal} {Nano Letters}\
  }\textbf {\bibinfo {volume} {10}},\ \bibinfo {pages} {3909} (\bibinfo {year}
  {2010})}\BibitemShut {NoStop}%
\bibitem [{\citenamefont {Dames}(2013)}]{Dames2013}%
  \BibitemOpen
  \bibfield  {author} {\bibinfo {author} {\bibfnamefont {C.}~\bibnamefont
  {Dames}},\ }\href {\doibase 10.1615/AnnualRevHeatTransfer.v16.20} {\bibfield
  {journal} {\bibinfo  {journal} {Annual Review of Heat Transfer}\ }\textbf
  {\bibinfo {volume} {16}},\ \bibinfo {pages} {7} (\bibinfo {year}
  {2013})}\BibitemShut {NoStop}%
\bibitem [{\citenamefont {Aubain}\ and\ \citenamefont
  {Bandaru}(2011)}]{Aubain2011}%
  \BibitemOpen
  \bibfield  {author} {\bibinfo {author} {\bibfnamefont {M.~S.}\ \bibnamefont
  {Aubain}}\ and\ \bibinfo {author} {\bibfnamefont {P.~R.}\ \bibnamefont
  {Bandaru}},\ }\href {\doibase 10.1063/1.3647318} {\bibfield  {journal}
  {\bibinfo  {journal} {Journal of Applied Physics}\ }\textbf {\bibinfo
  {volume} {110}},\ \bibinfo {pages} {084313} (\bibinfo {year}
  {2011})}\BibitemShut {NoStop}%
\bibitem [{\citenamefont {Bian}\ \emph {et~al.}(2003)\citenamefont {Bian},
  \citenamefont {Christofferson}, \citenamefont {Shakouri},\ and\ \citenamefont
  {Kozodoy}}]{Bian2003}%
  \BibitemOpen
  \bibfield  {author} {\bibinfo {author} {\bibfnamefont {Z.}~\bibnamefont
  {Bian}}, \bibinfo {author} {\bibfnamefont {J.}~\bibnamefont
  {Christofferson}}, \bibinfo {author} {\bibfnamefont {A.}~\bibnamefont
  {Shakouri}}, \ and\ \bibinfo {author} {\bibfnamefont {P.}~\bibnamefont
  {Kozodoy}},\ }\href {\doibase 10.1063/1.1623338} {\bibfield  {journal}
  {\bibinfo  {journal} {Applied Physics Letters}\ }\textbf {\bibinfo {volume}
  {83}},\ \bibinfo {pages} {3605} (\bibinfo {year} {2003})}\BibitemShut
  {NoStop}%
\bibitem [{\citenamefont {Zhang}\ \emph {et~al.}(2006)\citenamefont {Zhang},
  \citenamefont {Christofferson}, \citenamefont {Shakouri}, \citenamefont
  {{Deyu Li}}, \citenamefont {Majumdar}, \citenamefont {{Yiying Wu}},
  \citenamefont {{Rong Fan}},\ and\ \citenamefont {{Peidong
  Yang}}}]{Zhang2006}%
  \BibitemOpen
  \bibfield  {author} {\bibinfo {author} {\bibfnamefont {Y.}~\bibnamefont
  {Zhang}}, \bibinfo {author} {\bibfnamefont {J.}~\bibnamefont
  {Christofferson}}, \bibinfo {author} {\bibfnamefont {A.}~\bibnamefont
  {Shakouri}}, \bibinfo {author} {\bibnamefont {{Deyu Li}}}, \bibinfo {author}
  {\bibfnamefont {A.}~\bibnamefont {Majumdar}}, \bibinfo {author} {\bibnamefont
  {{Yiying Wu}}}, \bibinfo {author} {\bibnamefont {{Rong Fan}}}, \ and\
  \bibinfo {author} {\bibnamefont {{Peidong Yang}}},\ }\href {\doibase
  10.1109/TNANO.2005.861769} {\bibfield  {journal} {\bibinfo  {journal} {IEEE
  Transactions On Nanotechnology}\ }\textbf {\bibinfo {volume} {5}},\ \bibinfo
  {pages} {67} (\bibinfo {year} {2006})}\BibitemShut {NoStop}%
\bibitem [{\citenamefont {Favaloro}\ \emph {et~al.}(2014)\citenamefont
  {Favaloro}, \citenamefont {Ziabari}, \citenamefont {Bahk}, \citenamefont
  {Burke}, \citenamefont {Lu}, \citenamefont {Bowers}, \citenamefont {Gossard},
  \citenamefont {Bian},\ and\ \citenamefont {Shakouri}}]{Favaloro2014}%
  \BibitemOpen
  \bibfield  {author} {\bibinfo {author} {\bibfnamefont {T.}~\bibnamefont
  {Favaloro}}, \bibinfo {author} {\bibfnamefont {A.}~\bibnamefont {Ziabari}},
  \bibinfo {author} {\bibfnamefont {J.-H.}\ \bibnamefont {Bahk}}, \bibinfo
  {author} {\bibfnamefont {P.}~\bibnamefont {Burke}}, \bibinfo {author}
  {\bibfnamefont {H.}~\bibnamefont {Lu}}, \bibinfo {author} {\bibfnamefont
  {J.}~\bibnamefont {Bowers}}, \bibinfo {author} {\bibfnamefont
  {A.}~\bibnamefont {Gossard}}, \bibinfo {author} {\bibfnamefont
  {Z.}~\bibnamefont {Bian}}, \ and\ \bibinfo {author} {\bibfnamefont
  {A.}~\bibnamefont {Shakouri}},\ }\href {\doibase 10.1063/1.4885198}
  {\bibfield  {journal} {\bibinfo  {journal} {Journal of Applied Physics}\
  }\textbf {\bibinfo {volume} {116}},\ \bibinfo {pages} {034501} (\bibinfo
  {year} {2014})}\BibitemShut {NoStop}%
\bibitem [{\citenamefont {Maize}\ \emph {et~al.}(2014)\citenamefont {Maize},
  \citenamefont {Pavlidis}, \citenamefont {Heller}, \citenamefont {Yates},
  \citenamefont {Kendig}, \citenamefont {Graham},\ and\ \citenamefont
  {Shakouri}}]{Maize2014}%
  \BibitemOpen
  \bibfield  {author} {\bibinfo {author} {\bibfnamefont {K.}~\bibnamefont
  {Maize}}, \bibinfo {author} {\bibfnamefont {G.}~\bibnamefont {Pavlidis}},
  \bibinfo {author} {\bibfnamefont {E.}~\bibnamefont {Heller}}, \bibinfo
  {author} {\bibfnamefont {L.}~\bibnamefont {Yates}}, \bibinfo {author}
  {\bibfnamefont {D.}~\bibnamefont {Kendig}}, \bibinfo {author} {\bibfnamefont
  {S.}~\bibnamefont {Graham}}, \ and\ \bibinfo {author} {\bibfnamefont
  {A.}~\bibnamefont {Shakouri}},\ }in\ \href {\doibase
  10.1109/CSICS.2014.6978561} {\emph {\bibinfo {booktitle} {2014 IEEE Compound
  Semiconductor Integrated Circuit Symposium (CSICS)}}}\ (\bibinfo  {publisher}
  {IEEE},\ \bibinfo {year} {2014})\ pp.\ \bibinfo {pages} {1--8}\BibitemShut
  {NoStop}%
\bibitem [{\citenamefont {Nadri}\ \emph {et~al.}(2019)\citenamefont {Nadri},
  \citenamefont {Moore}, \citenamefont {Sauber}, \citenamefont {Xie},
  \citenamefont {Cyberey}, \citenamefont {Gaskins}, \citenamefont
  {Lichtenberger}, \citenamefont {{Scott Barker}}, \citenamefont {Hopkins},
  \citenamefont {Zebarjadi},\ and\ \citenamefont {Weikle}}]{Nadri2019}%
  \BibitemOpen
  \bibfield  {author} {\bibinfo {author} {\bibfnamefont {S.}~\bibnamefont
  {Nadri}}, \bibinfo {author} {\bibfnamefont {C.~M.}\ \bibnamefont {Moore}},
  \bibinfo {author} {\bibfnamefont {N.~D.}\ \bibnamefont {Sauber}}, \bibinfo
  {author} {\bibfnamefont {L.}~\bibnamefont {Xie}}, \bibinfo {author}
  {\bibfnamefont {M.~E.}\ \bibnamefont {Cyberey}}, \bibinfo {author}
  {\bibfnamefont {J.~T.}\ \bibnamefont {Gaskins}}, \bibinfo {author}
  {\bibfnamefont {A.~W.}\ \bibnamefont {Lichtenberger}}, \bibinfo {author}
  {\bibfnamefont {N.}~\bibnamefont {{Scott Barker}}}, \bibinfo {author}
  {\bibfnamefont {P.~E.}\ \bibnamefont {Hopkins}}, \bibinfo {author}
  {\bibfnamefont {M.}~\bibnamefont {Zebarjadi}}, \ and\ \bibinfo {author}
  {\bibfnamefont {R.~M.}\ \bibnamefont {Weikle}},\ }\href {\doibase
  10.1109/TED.2018.2880915} {\bibfield  {journal} {\bibinfo  {journal} {IEEE
  Transactions on Electron Devices}\ }\textbf {\bibinfo {volume} {66}},\
  \bibinfo {pages} {349} (\bibinfo {year} {2019})}\BibitemShut {NoStop}%
\bibitem [{\citenamefont {Yu}\ \emph {et~al.}(2010)\citenamefont {Yu},
  \citenamefont {Mitrovic}, \citenamefont {Tham}, \citenamefont {Varghese},\
  and\ \citenamefont {Heath}}]{Yu2010}%
  \BibitemOpen
  \bibfield  {author} {\bibinfo {author} {\bibfnamefont {J.-K.}\ \bibnamefont
  {Yu}}, \bibinfo {author} {\bibfnamefont {S.}~\bibnamefont {Mitrovic}},
  \bibinfo {author} {\bibfnamefont {D.}~\bibnamefont {Tham}}, \bibinfo {author}
  {\bibfnamefont {J.}~\bibnamefont {Varghese}}, \ and\ \bibinfo {author}
  {\bibfnamefont {J.~R.}\ \bibnamefont {Heath}},\ }\href {\doibase
  10.1038/nnano.2010.149} {\bibfield  {journal} {\bibinfo  {journal} {Nature
  Nanotechnology}\ }\textbf {\bibinfo {volume} {5}},\ \bibinfo {pages} {718}
  (\bibinfo {year} {2010})}\BibitemShut {NoStop}%
\bibitem [{\citenamefont {Hao}\ \emph {et~al.}(2016)\citenamefont {Hao},
  \citenamefont {Xiao},\ and\ \citenamefont {Zhao}}]{Hao2016}%
  \BibitemOpen
  \bibfield  {author} {\bibinfo {author} {\bibfnamefont {Q.}~\bibnamefont
  {Hao}}, \bibinfo {author} {\bibfnamefont {Y.}~\bibnamefont {Xiao}}, \ and\
  \bibinfo {author} {\bibfnamefont {H.}~\bibnamefont {Zhao}},\ }\href {\doibase
  10.1063/1.4959984} {\bibfield  {journal} {\bibinfo  {journal} {Journal of
  Applied Physics}\ }\textbf {\bibinfo {volume} {120}},\ \bibinfo {pages}
  {065101} (\bibinfo {year} {2016})}\BibitemShut {NoStop}%
\bibitem [{\citenamefont {Tang}\ \emph {et~al.}(2010)\citenamefont {Tang},
  \citenamefont {Wang}, \citenamefont {Lee}, \citenamefont {Fardy},
  \citenamefont {Huo}, \citenamefont {Russell},\ and\ \citenamefont
  {Yang}}]{Tang2010}%
  \BibitemOpen
  \bibfield  {author} {\bibinfo {author} {\bibfnamefont {J.}~\bibnamefont
  {Tang}}, \bibinfo {author} {\bibfnamefont {H.-T.}\ \bibnamefont {Wang}},
  \bibinfo {author} {\bibfnamefont {D.~H.}\ \bibnamefont {Lee}}, \bibinfo
  {author} {\bibfnamefont {M.}~\bibnamefont {Fardy}}, \bibinfo {author}
  {\bibfnamefont {Z.}~\bibnamefont {Huo}}, \bibinfo {author} {\bibfnamefont
  {T.~P.}\ \bibnamefont {Russell}}, \ and\ \bibinfo {author} {\bibfnamefont
  {P.}~\bibnamefont {Yang}},\ }\href {\doibase 10.1021/nl102931z} {\bibfield
  {journal} {\bibinfo  {journal} {Nano Letters}\ }\textbf {\bibinfo {volume}
  {10}},\ \bibinfo {pages} {4279} (\bibinfo {year} {2010})}\BibitemShut
  {NoStop}%
\bibitem [{\citenamefont {Lim}\ \emph {et~al.}(2016)\citenamefont {Lim},
  \citenamefont {Wang}, \citenamefont {Tang}, \citenamefont {Andrews},
  \citenamefont {So}, \citenamefont {Lee}, \citenamefont {Lee}, \citenamefont
  {Russell},\ and\ \citenamefont {Yang}}]{Lim2016}%
  \BibitemOpen
  \bibfield  {author} {\bibinfo {author} {\bibfnamefont {J.}~\bibnamefont
  {Lim}}, \bibinfo {author} {\bibfnamefont {H.-t.}\ \bibnamefont {Wang}},
  \bibinfo {author} {\bibfnamefont {J.}~\bibnamefont {Tang}}, \bibinfo {author}
  {\bibfnamefont {S.~C.}\ \bibnamefont {Andrews}}, \bibinfo {author}
  {\bibfnamefont {H.}~\bibnamefont {So}}, \bibinfo {author} {\bibfnamefont
  {J.}~\bibnamefont {Lee}}, \bibinfo {author} {\bibfnamefont {D.~H.}\
  \bibnamefont {Lee}}, \bibinfo {author} {\bibfnamefont {T.~P.}\ \bibnamefont
  {Russell}}, \ and\ \bibinfo {author} {\bibfnamefont {P.}~\bibnamefont
  {Yang}},\ }\href {\doibase 10.1021/acsnano.5b05385} {\bibfield  {journal}
  {\bibinfo  {journal} {ACS Nano}\ }\textbf {\bibinfo {volume} {10}},\ \bibinfo
  {pages} {124} (\bibinfo {year} {2016})}\BibitemShut {NoStop}%
\bibitem [{\citenamefont {Nomura}\ \emph {et~al.}(2018)\citenamefont {Nomura},
  \citenamefont {Shiomi}, \citenamefont {Shiga},\ and\ \citenamefont
  {Anufriev}}]{Nomura2018}%
  \BibitemOpen
  \bibfield  {author} {\bibinfo {author} {\bibfnamefont {M.}~\bibnamefont
  {Nomura}}, \bibinfo {author} {\bibfnamefont {J.}~\bibnamefont {Shiomi}},
  \bibinfo {author} {\bibfnamefont {T.}~\bibnamefont {Shiga}}, \ and\ \bibinfo
  {author} {\bibfnamefont {R.}~\bibnamefont {Anufriev}},\ }\href {\doibase
  10.7567/JJAP.57.080101} {\bibfield  {journal} {\bibinfo  {journal} {Japanese
  Journal of Applied Physics}\ }\textbf {\bibinfo {volume} {57}},\ \bibinfo
  {pages} {080101} (\bibinfo {year} {2018})}\BibitemShut {NoStop}%
\bibitem [{\citenamefont {Christofferson}\ and\ \citenamefont
  {Shakouri}(2005)}]{Christofferson2005}%
  \BibitemOpen
  \bibfield  {author} {\bibinfo {author} {\bibfnamefont {J.}~\bibnamefont
  {Christofferson}}\ and\ \bibinfo {author} {\bibfnamefont {A.}~\bibnamefont
  {Shakouri}},\ }\href {\doibase 10.1063/1.1850632} {\bibfield  {journal}
  {\bibinfo  {journal} {Review of Scientific Instruments}\ }\textbf {\bibinfo
  {volume} {76}},\ \bibinfo {pages} {024903} (\bibinfo {year}
  {2005})}\BibitemShut {NoStop}%
\bibitem [{\citenamefont {Incropera}\ \emph {et~al.}(2007)\citenamefont
  {Incropera}, \citenamefont {DeWitt}, \citenamefont {Bergman},\ and\
  \citenamefont {Lavine}}]{Incropera2007}%
  \BibitemOpen
  \bibfield  {author} {\bibinfo {author} {\bibfnamefont {F.~P.}\ \bibnamefont
  {Incropera}}, \bibinfo {author} {\bibfnamefont {D.~P.}\ \bibnamefont
  {DeWitt}}, \bibinfo {author} {\bibfnamefont {T.~L.}\ \bibnamefont {Bergman}},
  \ and\ \bibinfo {author} {\bibfnamefont {A.~S.}\ \bibnamefont {Lavine}},\
  }\href@noop {} {\emph {\bibinfo {title} {{Fundamentals of Heat and Mass
  Transfer}}}},\ \bibinfo {edition} {6th}\ ed.\ (\bibinfo  {publisher}
  {Wiley},\ \bibinfo {address} {Hoboken, NJ},\ \bibinfo {year}
  {2007})\BibitemShut {NoStop}%
\bibitem [{\citenamefont {Asheghi}\ \emph {et~al.}(1998)\citenamefont
  {Asheghi}, \citenamefont {Touzelbaev}, \citenamefont {Goodson}, \citenamefont
  {Leung},\ and\ \citenamefont {Wong}}]{Asheghi1998}%
  \BibitemOpen
  \bibfield  {author} {\bibinfo {author} {\bibfnamefont {M.}~\bibnamefont
  {Asheghi}}, \bibinfo {author} {\bibfnamefont {M.~N.}\ \bibnamefont
  {Touzelbaev}}, \bibinfo {author} {\bibfnamefont {K.~E.}\ \bibnamefont
  {Goodson}}, \bibinfo {author} {\bibfnamefont {Y.~K.}\ \bibnamefont {Leung}},
  \ and\ \bibinfo {author} {\bibfnamefont {S.~S.}\ \bibnamefont {Wong}},\
  }\href {\doibase 10.1115/1.2830059} {\bibfield  {journal} {\bibinfo
  {journal} {Journal of Heat Transfer}\ }\textbf {\bibinfo {volume} {120}},\
  \bibinfo {pages} {30} (\bibinfo {year} {1998})}\BibitemShut {NoStop}%
\bibitem [{\citenamefont {Kang}\ \emph {et~al.}(2008)\citenamefont {Kang},
  \citenamefont {Koh}, \citenamefont {Chiritescu}, \citenamefont {Zheng},\ and\
  \citenamefont {Cahill}}]{Kang2008}%
  \BibitemOpen
  \bibfield  {author} {\bibinfo {author} {\bibfnamefont {K.}~\bibnamefont
  {Kang}}, \bibinfo {author} {\bibfnamefont {Y.~K.}\ \bibnamefont {Koh}},
  \bibinfo {author} {\bibfnamefont {C.}~\bibnamefont {Chiritescu}}, \bibinfo
  {author} {\bibfnamefont {X.}~\bibnamefont {Zheng}}, \ and\ \bibinfo {author}
  {\bibfnamefont {D.~G.}\ \bibnamefont {Cahill}},\ }\href {\doibase
  10.1063/1.3020759} {\bibfield  {journal} {\bibinfo  {journal} {Review of
  Scientific Instruments}\ }\textbf {\bibinfo {volume} {79}},\ \bibinfo {pages}
  {114901} (\bibinfo {year} {2008})}\BibitemShut {NoStop}%
\bibitem [{\citenamefont {Hopkins}\ \emph {et~al.}(2010)\citenamefont
  {Hopkins}, \citenamefont {Serrano}, \citenamefont {Phinney}, \citenamefont
  {Kearney}, \citenamefont {Grasser},\ and\ \citenamefont
  {Harris}}]{Hopkins2010}%
  \BibitemOpen
  \bibfield  {author} {\bibinfo {author} {\bibfnamefont {P.~E.}\ \bibnamefont
  {Hopkins}}, \bibinfo {author} {\bibfnamefont {J.~R.}\ \bibnamefont
  {Serrano}}, \bibinfo {author} {\bibfnamefont {L.~M.}\ \bibnamefont
  {Phinney}}, \bibinfo {author} {\bibfnamefont {S.~P.}\ \bibnamefont
  {Kearney}}, \bibinfo {author} {\bibfnamefont {T.~W.}\ \bibnamefont
  {Grasser}}, \ and\ \bibinfo {author} {\bibfnamefont {C.~T.}\ \bibnamefont
  {Harris}},\ }\href {\doibase 10.1115/1.4000993} {\bibfield  {journal}
  {\bibinfo  {journal} {Journal of Heat Transfer}\ }\textbf {\bibinfo {volume}
  {132}} (\bibinfo {year} {2010}),\ 10.1115/1.4000993}\BibitemShut {NoStop}%
\bibitem [{\citenamefont {Schmidt}\ \emph {et~al.}(2008)\citenamefont
  {Schmidt}, \citenamefont {Chen},\ and\ \citenamefont {Chen}}]{Schmidt2008}%
  \BibitemOpen
  \bibfield  {author} {\bibinfo {author} {\bibfnamefont {A.~J.}\ \bibnamefont
  {Schmidt}}, \bibinfo {author} {\bibfnamefont {X.}~\bibnamefont {Chen}}, \
  and\ \bibinfo {author} {\bibfnamefont {G.}~\bibnamefont {Chen}},\ }\href
  {\doibase 10.1063/1.3006335} {\bibfield  {journal} {\bibinfo  {journal}
  {Review of Scientific Instruments}\ }\textbf {\bibinfo {volume} {79}},\
  \bibinfo {pages} {114902} (\bibinfo {year} {2008})}\BibitemShut {NoStop}%
\bibitem [{\citenamefont {Jiang}\ \emph {et~al.}(2017)\citenamefont {Jiang},
  \citenamefont {Qian}, \citenamefont {Gu},\ and\ \citenamefont
  {Yang}}]{Jiang2017}%
  \BibitemOpen
  \bibfield  {author} {\bibinfo {author} {\bibfnamefont {P.}~\bibnamefont
  {Jiang}}, \bibinfo {author} {\bibfnamefont {X.}~\bibnamefont {Qian}},
  \bibinfo {author} {\bibfnamefont {X.}~\bibnamefont {Gu}}, \ and\ \bibinfo
  {author} {\bibfnamefont {R.}~\bibnamefont {Yang}},\ }\href {\doibase
  10.1002/adma.201701068} {\bibfield  {journal} {\bibinfo  {journal} {Advanced
  Materials}\ }\textbf {\bibinfo {volume} {29}},\ \bibinfo {pages} {1701068}
  (\bibinfo {year} {2017})}\BibitemShut {NoStop}%
\bibitem [{\citenamefont {Lide}(2005)}]{Lide2005}%
  \BibitemOpen
  \bibinfo {editor} {\bibfnamefont {D.~R.}\ \bibnamefont {Lide}},\ ed.,\
  \href@noop {} {\emph {\bibinfo {title} {{CRC Handbook of Chemistry and
  Physics}}}}\ (\bibinfo  {publisher} {CRC Press LLC},\ \bibinfo {year}
  {2005})\BibitemShut {NoStop}%
\bibitem [{\citenamefont {Favaloro}\ \emph {et~al.}(2015)\citenamefont
  {Favaloro}, \citenamefont {Bahk},\ and\ \citenamefont
  {Shakouri}}]{Favaloro2015}%
  \BibitemOpen
  \bibfield  {author} {\bibinfo {author} {\bibfnamefont {T.}~\bibnamefont
  {Favaloro}}, \bibinfo {author} {\bibfnamefont {J.-H.}\ \bibnamefont {Bahk}},
  \ and\ \bibinfo {author} {\bibfnamefont {A.}~\bibnamefont {Shakouri}},\
  }\href {\doibase 10.1063/1.4907354} {\bibfield  {journal} {\bibinfo
  {journal} {Review of Scientific Instruments}\ }\textbf {\bibinfo {volume}
  {86}},\ \bibinfo {pages} {024903} (\bibinfo {year} {2015})}\BibitemShut
  {NoStop}%
\bibitem [{\citenamefont {Touloukian}\ \emph {et~al.}(1970)\citenamefont
  {Touloukian}, \citenamefont {Powel}, \citenamefont {Ho},\ and\ \citenamefont
  {Klemens}}]{Touloukian1970}%
  \BibitemOpen
  \bibfield  {author} {\bibinfo {author} {\bibfnamefont {Y.~S.}\ \bibnamefont
  {Touloukian}}, \bibinfo {author} {\bibfnamefont {R.~W.}\ \bibnamefont
  {Powel}}, \bibinfo {author} {\bibfnamefont {C.~Y.}\ \bibnamefont {Ho}}, \
  and\ \bibinfo {author} {\bibfnamefont {P.~G.}\ \bibnamefont {Klemens}},\ }in\
  \href@noop {} {\emph {\bibinfo {booktitle} {Thermophysical Properties of
  Matter}}}\ (\bibinfo {year} {1970})\BibitemShut {NoStop}%
\bibitem [{\citenamefont {Mahajan}\ \emph {et~al.}(2011)\citenamefont
  {Mahajan}, \citenamefont {Subbarayan},\ and\ \citenamefont
  {Sammakia}}]{Mahajan2011}%
  \BibitemOpen
  \bibfield  {author} {\bibinfo {author} {\bibfnamefont {S.~S.}\ \bibnamefont
  {Mahajan}}, \bibinfo {author} {\bibfnamefont {G.}~\bibnamefont {Subbarayan}},
  \ and\ \bibinfo {author} {\bibfnamefont {B.~G.}\ \bibnamefont {Sammakia}},\
  }\href {\doibase 10.1109/TCPMT.2011.2112356} {\bibfield  {journal} {\bibinfo
  {journal} {IEEE Transactions on Components, Packaging and Manufacturing
  Technology}\ }\textbf {\bibinfo {volume} {1}},\ \bibinfo {pages} {1132}
  (\bibinfo {year} {2011})}\BibitemShut {NoStop}%
\bibitem [{\citenamefont {Chen}\ \emph {et~al.}(2012)\citenamefont {Chen},
  \citenamefont {Zhang},\ and\ \citenamefont {Li}}]{Chen2012}%
  \BibitemOpen
  \bibfield  {author} {\bibinfo {author} {\bibfnamefont {J.}~\bibnamefont
  {Chen}}, \bibinfo {author} {\bibfnamefont {G.}~\bibnamefont {Zhang}}, \ and\
  \bibinfo {author} {\bibfnamefont {B.}~\bibnamefont {Li}},\ }\href {\doibase
  10.1063/1.4754513} {\bibfield  {journal} {\bibinfo  {journal} {Journal of
  Applied Physics}\ }\textbf {\bibinfo {volume} {112}},\ \bibinfo {pages}
  {064319} (\bibinfo {year} {2012})}\BibitemShut {NoStop}%
\bibitem [{\citenamefont {Lampin}\ \emph {et~al.}(2012)\citenamefont {Lampin},
  \citenamefont {Nguyen}, \citenamefont {Francioso},\ and\ \citenamefont
  {Cleri}}]{Lampin2012}%
  \BibitemOpen
  \bibfield  {author} {\bibinfo {author} {\bibfnamefont {E.}~\bibnamefont
  {Lampin}}, \bibinfo {author} {\bibfnamefont {Q.-H.}\ \bibnamefont {Nguyen}},
  \bibinfo {author} {\bibfnamefont {P.~A.}\ \bibnamefont {Francioso}}, \ and\
  \bibinfo {author} {\bibfnamefont {F.}~\bibnamefont {Cleri}},\ }\href
  {\doibase 10.1063/1.3698325} {\bibfield  {journal} {\bibinfo  {journal}
  {Applied Physics Letters}\ }\textbf {\bibinfo {volume} {100}},\ \bibinfo
  {pages} {131906} (\bibinfo {year} {2012})}\BibitemShut {NoStop}%
\bibitem [{\citenamefont {Liu}\ and\ \citenamefont {Asheghi}(2006)}]{Liu2006}%
  \BibitemOpen
  \bibfield  {author} {\bibinfo {author} {\bibfnamefont {W.}~\bibnamefont
  {Liu}}\ and\ \bibinfo {author} {\bibfnamefont {M.}~\bibnamefont {Asheghi}},\
  }\href {\doibase 10.1115/1.2130403} {\bibfield  {journal} {\bibinfo
  {journal} {Journal of Heat Transfer}\ }\textbf {\bibinfo {volume} {128}},\
  \bibinfo {pages} {75} (\bibinfo {year} {2006})}\BibitemShut {NoStop}%
\end{thebibliography}%

\end{document}